\documentclass
[preprint,showpacs,byrevtex,10pt,twocolumn,tightenlines,prl,letterpaper]{revtex4}%
\usepackage{amsfonts}
\usepackage{amsmath}
\usepackage{amssymb}
\usepackage[dvips]{graphicx}
\usepackage{hyphenat}%
\providecommand{\U}[1]{\protect\rule{.1in}{.1in}}

\begin{document}
\preprint{ }
\title{Slow Dynamics of the Electron-Glasses; the Role of Disorder}
\author{Z. Ovadyahu}
\affiliation{Racah Institute of Physics, The Hebrew University, Jerusalem 91904, Israel }


\begin{abstract}
We examine in this work the role of disorder in contributing to the sluggish
relaxation observed in intrinsic electron-glasses. Our approach is guided by
several empirical observations: First and foremost, Anderson localization is a
pre-requisite for observing these nonequilibrium phenomena. Secondly, sluggish
relaxation appears to favor Anderson-insulators with relatively large
Fermi-energies (hence proportionally large disorder). These observations
motivated us to consider a way to measure the underlying disorder in a
realistic Anderson insulator. Optical study using a series of amorphous
indium-oxide (In$_{\text{x}}$O) establish a simple connection between
carrier-concentration and the disorder necessary to approach the
metal-insulator transition from the insulating side. This is used to estimate
the typical magnitude of the quenched potential-fluctuation in the
electron-glass phase of this system. The implications of our findings on the
slow dynamics of Anderson-insulators are discussed. In particular, the reason
for the absence of a memory-dip and the accompanying electron-glass effects in
lightly-doped semiconductors emerges as a natural consequence of their weak disorder.

\end{abstract}
\maketitle

\section{Introduction}

Theoretical considerations anticipating nonequilibrium effects in
Anderson-localized systems were described in a number of papers
\cite{1,2,3,4,5,6,7,8,9,10,11,12}. These were based on the interplay between
disorder and Coulomb interactions leading to an electron-glass phase.

Over the last few decades there were several experimental studies that seem to
give support to these expectations \cite{13,14,15}. The dynamics that
characterizes the approach to equilibrium of these systems is sluggish;
relaxation of the excess conductance produced by driving the system far from
the equilibrium, was observed to persist for many hours in some cases
\cite{16}. The long relaxation of the electronic system makes it possible to
observe a modulation of the \textit{single-particle} density-of-states (DOS)
in field-effect experiments \cite{17}. This feature, called a `memory-dip'
(MD), appears as a cusp-like minimum in the conductance versus gate-voltage.
An example is illustrated in the inset to Fig.1. The memory-dip, presumably
\cite{5,9} a reflection of an underlying Coulomb-gap \cite{20,21}, is the
identifying feature of intrinsic electron-glass \cite{5,6,9,11,22}.

To date, a memory-dip has been observed in seven different Anderson insulators
(listed in Fig.1). The figure shows an empirical correlation between the
typical width $\Gamma$ of the MD and the carrier-concentration \textit{N} of
the material. The list of Anderson insulators that exhibit electron-glass
properties include all types of degenerate Fermi systems; n-type
semiconductors (Tl$_{\text{2}}$O$_{\text{3-x}}$, In$_{\text{x}}$O,
In$_{\text{2}}$O$_{\text{3-x}}$), p-type semiconductors (GeSb$_{\text{x}}%
$Te$_{\text{y}}$, GeTe), and a metal (Be).

The systematic dependence of the MD characteristic width on
carrier-concentration, an \textit{electronic} property, is consistent with the
expectation that the phenomenon is intrinsic. Due to lack of screening in the
Anderson-insulating phase, Coulomb interaction may be comparable in magnitude
to the quenched disorder, and therefore these competing ingredients
responsible for glassy behavior are always present in Anderson insulators.
\begin{figure}[ptb]%
\centering
\includegraphics[
height=2.3341in,
width=3.0381in
]%
{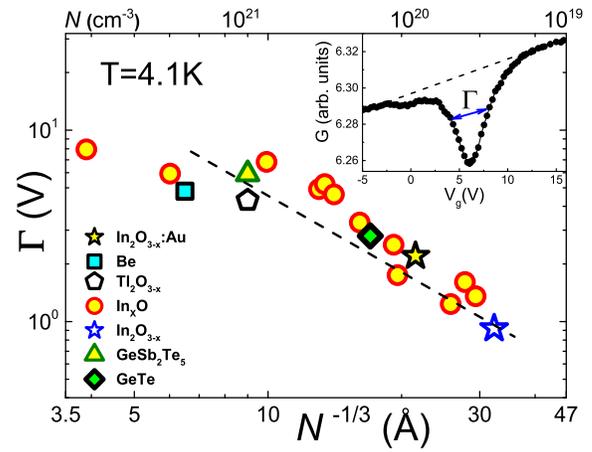}%
\caption{(color online) The typical width of the memory-dip $\Gamma$ (defined
in the inset) as function of carrier-concentration \textit{N} for several
Anderson-insulators. \ Data are taken from \cite{18} and \cite{19}. Inset:
Field-effect measurement for amorphous indium-oxide film with \textit{N}%
$\approx$3x10$^{\text{20}}$cm$^{\text{-3}}$ revealing a memory-dip centered at
+6V, which is the gate-voltage maintained between the sample and the gate for
24 hours before sweeping the gate-voltage over the range shown. The
In$_{\text{x}}$O was separated from the gate by 0.5$\mu$m of SiO$_{\text{2}}$
(the $\Gamma$ versus \textit{N }results in the main figure are all normalized
to this geometry). The dashed line depicts the law: $\Gamma\propto$
\textit{N}$^{\text{-1/3}}$ which is consistent with the Coulomb-gap behavior
suggested by the models in \cite{5} and \cite{9}.}%
\end{figure}

There seem to be more requirements on the material to allow observation of
electron-glass effects. Note that all the system$\;$in Fig.1 are made from
materials with carrier-concentration \textit{N} limited to a range
10$^{\text{22}}$cm$^{\text{-3}}$%
$>$%
\textit{N}$\geq$5x10$^{\text{19}}$cm$^{\text{-3}}$. The absence of systems
with \textit{N}%
$>$%
10$^{\text{22}}$cm$^{\text{-3}}$ from this list is not surprising; a
pre-requisite for observing electron-glass effects is Anderson-localization,
which is hard to achieve in a system with large \textit{N }unless by making it
granular (on which we remark later). Less clear is the limit of low
carrier-concentration. No memory-dip has been reported in a system with
carrier-concentration smaller than $\approx$10$^{\text{19}}$cm$^{\text{-3}}$
in any Anderson-localized system like Si or GaAs. Two-dimensional samples of
these materials may be tuned to show insulating behavior and were extensively
studied in the hopping regime. Their near-equilibrium transport properties
(like conductivity versus temperature) are not qualitatively different than
those of the systems in Fig.1. The absence of a MD in these systems has been a
vexing question for quite some time.

It was conjectured \cite{22} that relaxation processes in lightly-doped
semiconductors (with \textit{N} $\leq$10$^{\text{17}}$cm$^{\text{-3}}$) are
too fast to allow their signature to be captured by field-effect measurements.
This was inspired by the observation of relaxation dynamics in a series of
amorphous indium-oxide (In$_{\text{x}}$O) films \cite{23,22}. These can be
fabricated with different \textit{N} values, covering \textit{N}$\approx
$5x10$^{\text{18}}$cm$^{\text{-3}}$ to $\hspace{0.01in}$\textit{N}$\approx
$5x10$^{\text{21}}$cm$^{\text{-3}}$ by controlling the In-O ratio \cite{24}%
$.$The typical relaxation times of these films sharply diminished \cite{23}
once the carrier-concentration was reduced below \textit{N}$\approx
$10$^{\text{20}}$cm$^{\text{-3}}$. Note that, given the sample-gate
capacitance (as well as the other parasitic circuit capacitances), the
field-effect temporal resolution is severely limited for high resistance
samples. Consistent with this conjecture, ultrafast relaxation in a
lightly-doped semiconductor was reported in phosphorous-doped Si \cite{25}.

What needs to be clarified is the role of carrier-concentration in affecting
the dynamics of electron-glasses.

A phenomenon that is highly sensitive to the value of \textit{N} is often
indicative of a many-body mechanism playing a role. Mechanisms that were
considered in this regard include correlated many-particle transitions
\cite{26}, and the Anderson orthogonality catastrophe \cite{27}. A model by
Leggett at al, based on coupling to an electronic-bath \cite{28}, is
consistent with many of the observed features \cite{27}. These mechanisms, in
conjunction with quenched disorder and hierarchical constraints, are likely
contributing factors in the sluggish relaxation to some degree. However, to
account for relaxation times of the order of hours with these scenarios still
requires that the \textit{bare} tunneling probabilities are much smaller than
those involved in the dc conductivity.

The purpose of this work is to investigate the role played by disorder in
slowing down charge rearrangement processes involved in the energy relaxation
of electron-glasses. This is motivated by an alternative interpretation of the
decisive role that seems to be played by the carrier-concentration of the
system. Note that, at the transition, the magnitude of disorder in an Anderson
insulator with large carrier-concentration must be larger than that of a
system with low carrier-concentration, which may be summarized by a simple
relation between the Fermi energy of the system and the critical disorder. It
will be shown in this paper that this remains true deep into the insulating
regime. Therefore, the empirical observation that the relaxation becomes
slower once N becomes larger, may turn out to be related to the weaker
disorder rather than to various many--body effects (although the latter may be
quite effective in further slowing down the relaxation).

The first step in examining this conjecture calls for assessing the magnitude
of the disorder in the Anderson-insulating phase and its relation to the
carrier-concentration of the system. We describe a set of experiments on
several batches of amorphous indium-oxide with different
carrier-concentrations, and with different degree of disorder in each batch.
These data are analyzed to demonstrate a simple relation between disorder and
the Fermi-energy of a batch with a given N deep into the insulating regime.
This, augmented by further arguments, is used to explain why
Anderson-localized systems with low carrier-concentrations are unlikely to
exhibit intrinsic electron-glass effects with relaxation times longer than few seconds.

To understand the motivation for our approach to the problem, we first review
the main experimental findings pertinent to the electron-glass dynamics.

\subsection{Basic features of the dynamics in glassy Anderson-insulators}

Due to the lack of a concrete time-scale in their temporal relaxation
(generally, a power-law), dynamics of glasses cannot be uniquely quantified.
Tests to study, on a \textit{relative} basis, how various agents affect the
dynamics were performed on Anderson insulating crystalline indium-oxide,
In$_{\text{2}}$O$_{\text{3-x}}$ and on In$_{\text{x}}$O with various
compositions \cite{23,27,29}. These studies were also limited to effectively
two-dimensional (2D) samples with sheet resistances R$_{\square}$ in the
range\ of $\approx$1M$\Omega$ to $\approx$500M$\Omega$ where the signal to
noise of the glassy features is favorable. With these system however, it was
possible to test dynamics over a large range of lateral dimensions, from
2$\mu$m to 10mm \cite{30}.

The first observation, already alluded to above, is that dynamics seem to
become faster when the carrier-concentration is smaller \cite{23}. Secondly,
all features associated with slow dynamics disappear as the system crosses
over to the diffusive regime \cite{22}.

Over the temperature range $\approx$ 2-8K there was no indication of dynamics
freezeout \cite{27,29}. This is in stark contrast with the behavior of
classical glasses where dynamics is quickly frozen below a certain temperature
\cite{31}. The sample conductance over this temperature range typically
changes by 1-2 orders of magnitude.

Transition-rates associated with conductivity of Anderson-insulators can be
expressed as: $\omega\exp[-$r/$\xi]$ were r is the hopping length and $\xi$
the localization-length These rates typically range between $\approx
$10$^{\text{10}}$sec$^{\text{-1}}$ to $\approx$10$^{\text{5}}$sec$^{\text{-1
}}$based on the transition probability through a bottle-neck resistor with
r/$\xi$ $\approx$6-15 (a typical value for the hopping regime) and assuming
attempt frequency $\omega$ of the order of 10$^{\text{12}}$sec$^{\text{-1}}$
which is commonly used as the prefactor in hopping conductivity. Here r is the
hopping-length and $\xi$ the localization-length. Relaxation rates of the
studied electron-glasses are obviously slower by many orders of magnitude.

Another demonstration that conductance and relaxation processes appear to be
different was recently observed in photoconductivity experiments on
GeSb$_{\text{x}}$Te$_{\text{y}}$ films in their glassy regime; Adding charge
to the system by optical excitation enhanced the conductance but it
\textit{slowed-down} the relaxation dynamics \cite{32}.

These observations suggest that relaxation and conductivity involve different
processes. It is natural then to consider \textit{extrinsic} effects as a
viable mechanism for the slow relaxation exhibited by these systems. There are
several candidates to choose from; ion-motion, surface-traps, grain-boundaries
\cite{33}, and negative-U centers \cite{34} are all potential sources for slow
dynamics. Their coupling to charge-carriers might be the reason for the
observed conductance relaxation observed in the experiments. In the first
place, this would immediately explain why relaxation from an
out-of-equilibrium state is controlled by different processes than those
involved in the dc conductance. It is also appealing as a source of a very
slow phenomenon, intimately connected with the ubiquitous 1/f-noise phenomenon
\cite{35}.

This however, is one of the problems with extrinsic mechanisms; flicker noise
is indeed ubiquitous, intrinsic electron-glassiness is \textit{not}. 1/f-noise
may be observed in the metallic as well as in the localized transport regime
while the electron-glass is an \textit{exclusive property of the localized
phase}. Furthermore, as mentioned above, a memory-dip, which is the
identifying feature of the intrinsic electron-glass, has not been seen in
\textit{any} lightly-doped semiconductors while 1/f noise is quite evident in
all these systems. GaAs samples in particular, exhibit prominent 1/f noise yet
no slow conductance relaxation that might have compromised their operation as
bolometers has been observed \cite{36}.

Other shortcomings of extrinsic scenarios include difficulties to account for
the dependence of the memory-dip on temperature and disorder \cite{22}, they
cannot explain the systematic dependence of the memory-dip width on
carrier-concentration (Fig.1), and they cannot account for the peculiar
evolution of the MD shape with temperature \cite{22,31}. Finally, the failure
of lightly-doped semiconductors to show a memory-dip remains enigmatic in this approach.

A more promising route to pursue is the `purely' electronic scenario which
accounts for most of the observed features \cite{5,9,10,11}, and as argued
below, the reason for the absence of intrinsic glass effects in low
carrier-concentration systems emerges naturally from this picture. In this
approach, both conductivity and relaxation in the electron-glass proceed via
charge-carriers transitions between localized states. The qualitative
difference between transitions involved in relaxation and those that control
conductivity is that conductance must include activated processes while
relaxation, in our scenario, is dominated by tunneling between states
differing in energy by $\delta\ll$ k$_{\text{B}}$T \cite{27}.

The driving force for the relaxation is minimizing the electrostatic energy
E$_{\text{el}}$=$\sum_{\text{i,j}}\frac{e^{\text{2}}}{\text{r}_{\text{i,j}%
}\varepsilon\text{(r}_{\text{i,j}}\text{)}}$ under the constraints set by the
disorder (assumed to be quenched). A simple example of an energy-reducing
event due to charge re-arrangement is illustrated in Fig.2.%
\begin{figure}[ptb]%
\centering
\includegraphics[
height=1.4676in,
width=3.039in
]%
{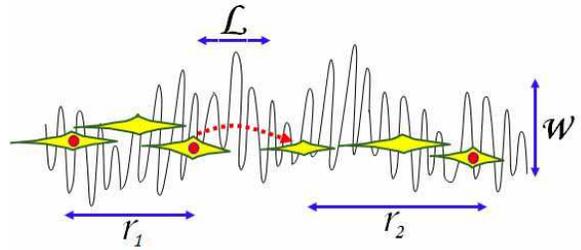}%
\caption{(color online) A schematic description of a proposed element of the
asymptotic energy-relaxation process. Charge-carriers (full circles) tunneling
through potential-fluctuations ($\mathcal{W}$ ) to lower the configurational
energy associated with the inter-particle Coulomb interaction.}%
\end{figure}

A transition of the type shown in Fig.2 for example, would result in an energy
release of the order of $\delta\varepsilon$=$\frac{\text{e}^{\text{2}%
}\text{(r}_{\text{2}}\text{-r}_{\text{1}}\text{)}}{\varepsilon\text{(r}%
_{\text{1,2}}\text{)r}_{\text{1}}\text{r}_{\text{2}}}$that amounts to a
substantial excess energy injected into the system (except for a rare
r$_{\text{2}}\approx$r$_{\text{1}}$ event). This may enhance the excess
conductivity via two possible mechanisms:

First, by generating an excess of nonequilibrium phonons. Although some of
these generated phonons would escape to the bath, it does not take much of
overheating to appreciably affect the conductivity in this regime. Consider
for example, a typical electron-glass with R$_{\square}\approx$10M$\Omega$ at
T=4K. This, when excited, will have an excess conductance $\Delta$G/G of the
order of $\approx$1\% (see, e/g., Fig.6 in \cite{37}). Overheating by
$\approx$5mK's (at 4K) is enough to generate this excess conductance given the
R(T) of such a sample (stretched exponential with a power of 1/3 or 1/2 with
activation energies of $\approx$5000K and 800K for the indium-oxide and Be
sample \cite{37} respectively). Even much less overheating will be needed
deeper in the electron glass regime; $\Delta$G/G increases algebraically with
disorder but the sensitivity of G to temperature increases exponentially with it.

Secondly, a transition event at any point R$_{\text{i}}$ in the system may
induce transitions in another site R$_{\text{j}}$ either directly via Coulomb
interaction (when
$\vert$%
R$_{\text{i}}$-R$_{\text{j}}$%
$\vert$%
$<$%
r$_{\text{h}}$) or indirectly via a succession of `avalanches' (for
$\vert$%
R$_{\text{i}}$-R$_{\text{j}}$%
$\vert$%
$>$%
r$_{\text{h}}$) \cite{38,39} where r$_{\text{h}}$ is the hopping-length. As
alluded to above, there is no metallic screening in the Anderson-localized
phase but at finite temperatures the Coulomb interaction is effectively cutoff
at distances larger than the hopping-length \cite{38,39}. These avalanches,
much like the Domino-Effect, ultimately spread to modulate the space-energy
configuration of the CCN thus affecting the measured conductance. Relaxation
of the excess-conductance would last as long as such energy-reducing
transitions occur. It should be noted that the probability of these
transitions to occur depends also on the electron-phonon coupling strength
(direct electron-electron inelastic transitions are essentially suppressed in
the strongly-localized system relative to its value in the diffusive regime
\cite{38}).

There is also a quantitative difference between transitions involved in the dc
conductance and those contributing to relaxation processes. Conductance is
determined by relatively fast transitions between pair of sites composing the
current-carrying-network (CCN). These are the pairs of sites connected by the
relatively high transition probabilities in the Miller-Abrahams conductance
distribution. In the hopping regime the CCN typically occupies a tiny fraction
of the system volume; most of the material is `dead-wood'; these are the high
resistance regions of the system \cite{40,41,42}, and this is naturally where
the slowest transitions occur. Energy-relaxation in the Anderson insulator
involves energy-reducing transitions anywhere in the system including in
particular the dead-wood regions.

The relative weight of the dead-wood gains in prominence when the disorder
gets larger or when the temperature gets smaller. In our conjecture, this
would account for the experimentally observed increase of the relative
magnitude of the MD resistance whether affected by disorder, field or
temperature \cite{43}.

The dead-wood, occupying the bulk of the system volume, holds most of the
excess-energy associated with the out-of-equilibrium state. Therefore,
conductance relaxation in this picture is essentially controlled by the
dead-wood slow-dynamics.

\subsection{Dynamics in the dead-wood}

The question we need to address concerns the transition rates of tunneling
events of the general type illustrated in Fig.2. Tunneling probability depends
exponentially on the distance $\mathcal{L}$ and on an effective-barrier V*.
Tunneling distance $\mathcal{L}$ for electrons could easily be much larger
than that typically found for ions and may extend over 10$^{\text{2}}%
$-10$^{\text{3}}$ atoms in solids. This balances out their transition rates as
compared with ions with their $\approx$10$^{\text{4}}$ times heavier-mass.
Tunneling probability may be vanishingly small if $\mathcal{L}$ is long
enough. However, the most likely transition-scale at the asymptotic regime of
the relaxation is perhaps of the order of the Bohr-radius a$_{\text{B}}$.
Transitions over longer distances likely proceed by series of short events.
Such `serial' events, as well as other `hierarchical processes', would further
slowdown the dynamics and should be treated separately.

The quantum transmission through a slab of an Anderson insulator of a finite
cross section and length has been worked out by Nikoli\'{c} and Dragomirova
\cite{44}. They studied the statistics of the associated eigenvalues as
function of disorder and for several lengths. The distribution of transmission
eigenvalues was found to be very wide. In the presence of moderate disorder it
included a finite portion of near resonance channels which may play a role in
local hierarchical processes but they do not directly contribute to
relaxation. On the other hand, the distribution of transmission probability
was heavily skewed in favor of low transmission channels \cite{43} even for
weak disorder while the resonances naturally disappeared when critical
disorder was imposed. It is plausible that including electron-electron
interaction in such calculations will only enhance the peak of the low
transition probabilities in the spectrum. Slow relaxation events should
therefore be abundant in the system.

Tunneling probability $\gamma$ through a simple (square) barrier may be
estimated by a WKB expression:%
\begin{equation}
\gamma\propto\exp\left(  \text{-2}\left[  \frac{\text{2m*V*}}{\hslash
^{\text{2}}}\right]  ^{\text{1/2}}\cdot\mathcal{L}\right)  ;\text{V*=V-E}%
\end{equation}
where m* is the effective-mass of the charge carrier, V the barrier-height and
E the particle energy (which will be taken as E$_{\text{F}}$). The spatial
form of $\mathcal{W}$(\textbf{r}) in a realistic Anderson insulator obviously
requires a more elaborate treatment than addressed by the WKB approximation
(probably a numerical work along the lines of \cite{42}). However, the
exponential dependence on m* $\mathcal{L}$\ and V* should still be reflected
in a more detailed treatment. This was recently demonstrated in a numerical
study extending the calculations of \cite{44}; in the localized regime, the
probability to find transmission trajectory with value smaller than a given
$\gamma$ is exponential with $\mathcal{W}$ $^{\text{1/2}}$ and $\mathcal{L}$
\cite{45}.

To estimate how slow a typical tunneling event may be, one needs to know the
magnitude of the V*. We shall assume it to be of order of the
potential-fluctuation $\mathcal{W}$. The value of $\mathcal{W}$ is the
distinguishing factor in determining whether electron-glass effects can be
observed by field-effect measurements in a given system. $\mathcal{W}$ may
vary by orders of magnitudes while a$_{\text{B}}$ is typically in the range of
20-50\AA . Larger values for a$_{\text{B}}$ are usually due to very light
effective-mass which counteracts the effect of longer tunneling-distance on
the tunneling probability.

To anticipate the discussion below, recall that a pre-requisite for
electron-glass behavior is Anderson-localization, which means that the
disorder energy $\mathcal{W}$ has to be comparable with, or larger than, the
kinetic energy E$_{\text{F}}$. All other things being equal, a system with
larger carrier-concentration must be more disordered to be Anderson-localized
and thus has larger $\mathcal{W}$. Lightly-doped semiconductors used for
hopping conductivity studies have typically \textit{N}$\leq$10$^{\text{17}}%
$cm$^{\text{-3}}$. Their associated Fermi-energy (and thus their $\mathcal{W}%
$) may be too small to sustain slow transitions. At comparable value of
resistivity and reduced-temperature k$_{\text{B}}$T/E$_{\text{F}}$, their
disorder is typically \textit{much} weaker than that of hopping systems with
\textit{N{}{}}$\geq$10$^{\text{20}}$cm$^{\text{-3}}$. What transpires from
these consideration is the need to know what actually is $\mathcal{W}$ for a
given Anderson-insulator. An attempt to deal with this elusive issue is our
next step.

\subsection{The `critical' disorder in Anderson-insulators}

There are several ways to characterize the magnitude of disorder in the
Anderson-localized phase. For example, the value of the localization length
$\xi$ is, in a way, a measure of disorder. However, determining $\xi$ involves
transport measurements at low temperatures where one probes just the CCN thus
ignoring the most disordered part of the sample. Another choice is the
Ioffe-Regel parameter k$_{\text{F}}\ell$ that may be estimated from Hall
effect and resistivity measurements at relatively high temperatures \cite{46}.
k$_{\text{F}}\ell$ may be taken as a measure of disorder whenever $\rho$ is
dominated by the elastic mean-free-path $\ell$. This condition is well-obeyed
in the vicinity of k$_{\text{F}}\ell\simeq$1 where the conductivity is
scale-independent \cite{47} and the estimate of k$_{\text{F}}\ell$ is then
least sensitive to the specific temperature at which $\rho$ is measured.
{\small I}n the regime k$_{\text{F}}\ell$%
$<$%
1 neither k$_{\text{F}}$nor $\ell$ have their usual meaning but k$_{\text{F}%
}\ell$ may still \ be a useful parameter to characterize disorder; it is just
a dimensionless parameter that decreases monotonically with disorder. On the
other hand, k$_{\text{F}}\ell$ is not simply related to static disorder once
k$_{\text{F}}\ell\lll$1 (or k$_{\text{F}}\ell\gg$1, a regime irrelevant for
this discussion anyhow). In these limiting cases the room-temperature
conductivity may be dominated by inelastic processes rather than by static disorder.

Given a system with a certain k$_{\text{F}}\ell$ one still requires a way to
assign a value for its $\mathcal{W}$. Theoretical estimates may be used for
the `critical disorder' $\mathcal{W}_{\mathcal{C}}$ the disorder necessary to
just localize the entire band. Estimates based on a non-interacting picture
yield the ratio $\mathcal{W}_{\mathcal{C}}$/I$\simeq$16.5 \cite{48}. Here I is
of the order of the band-width, typically several electron-volts. In these
models the Fermi-energy is taken at mid-band where the density-of-states is
highest. The models however are more vague when the Fermi-energy is near the
band-edge where $\partial$n$/\partial\mu$~may be rather small. Unfortunately,
this is invariably the situation in real Anderson-insulators \cite{48}. It is
certainly the case for each of the seven electron-glasses listed in Fig.1.

A more helpful approach then is to rely on the physics of the diffusive regime
and extrapolate to the transition point defined by the value of k$_{\text{F}%
}\ell$ at the transition to the localized phase obtainable from experiments. A
measure of disorder for a diffusive system is $\mathcal{W}$ =$\hslash$/$\tau$
where $\tau$ is the transport mean-free-time.

On the metallic side the ratio $\frac{\hslash/\tau}{E_{\text{F}}}=$ $\frac
{2}{k_{\text{F}}\ell}$, so at the metal-insulator transition:%
\begin{equation}
\mathcal{W}_{\mathcal{C}}=\beta\text{E}_{\text{F}}%
\end{equation}
where $\beta$ depends on the specific value of (k$_{\text{F}}\ell$%
)$_{\text{c}}$ - the Ioffe-Regel parameter evaluated at the transition to the
localized phase:%
\begin{equation}
\beta=\frac{2}{(k_{\text{F}}\ell)_{\text{c}}}%
\end{equation}

The experiments described next, suggest that the proportionality between
E$_{\text{F}}$ and $\mathcal{W}$ is still valid deep into the insulating
regime (k$_{\text{F}}\ell\ll$1), covering the entire range of resistances
probed in the electron-glass studies.

\subsection{Gauging disorder by monitoring optical properties; tuning disorder
in the amorphous indium-oxides}

A system that allows a continuous tuning of disorder for both sides of the
metal-insulator is amorphous indium-oxide In$_{\text{x}}$O. It may also be
prepared with a vastly different carrier-concentration (between $\approx
$5x10$^{\text{18}}$cm$^{\text{-3}}{}{}\ $to $\approx$6x10$^{\text{21}}%
$cm$^{\text{-3}}$) by controlling the In/O ratio. This makes it possible to
study the dependence of the electron-glass properties on carrier-concentration
and disorder \cite{24}, as well as the metal-insulator \cite{24} and the
superconductor-insulator transition \cite{49}.

The feasibility of tuning the system resistivity by heat-treatment is a rather
unique trait of In$_{\text{x}}$O. It is possible to vary the room-temperature
resistance of the as-prepared In$_{\text{x}}$O sample by up to 5 orders of
magnitude while maintaining its amorphous structure and composition (and hence
carrier-concentration). A before-and-after diffraction pattern illustrating
the preservation of amorphicity in the heat-treatment is shown in Fig.3.%
\begin{figure}[ptb]%
\centering
\includegraphics[
height=4.8456in,
width=3.039in
]%
{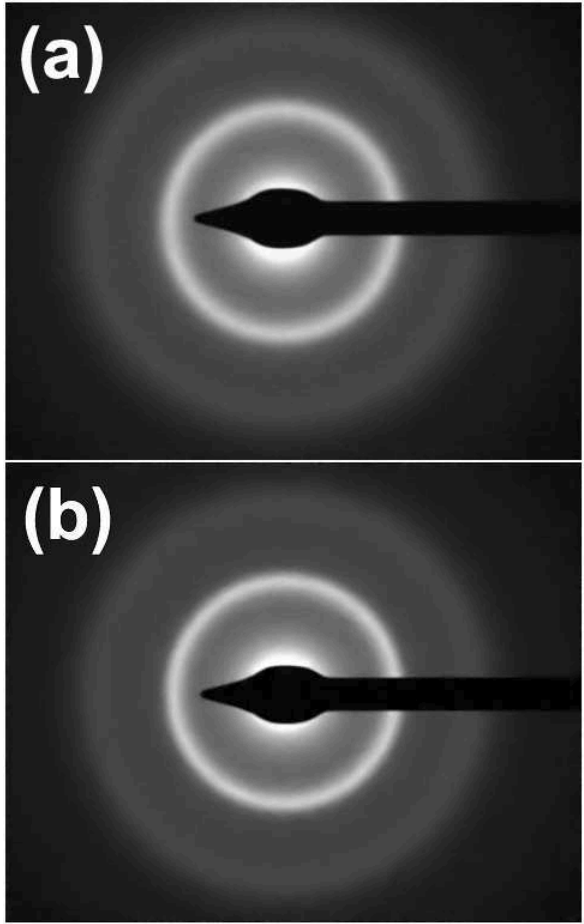}%
\caption{Electron diffraction patterns (using 200kV beam) of a $\approx$50 nm
In$_{\text{x}}$O film for the as-prepared specimen with sheet resistance
R$_{\square}$ $>$10$^{\text{8}}\Omega$ (a) and after thermal annealing for 34
days yielding R$_{\square}$ $\approx$7x10$^{\text{4}}\Omega$ (b). The general
appearance of the pattern seem unchanged in the annealing but a careful
comparison of the strong diffraction ring between the (a) and (b) micrographs
reveals slightly less fuzziness in the annealed sample (illustrated in Fig.4
below).}%
\end{figure}
This figure demonstrates that the amorphous structure remains intact during
thermal annealing. Actually, one is hard pressed to see a difference in the
before-and-after diffraction patterns (Fig.3). It takes a careful measurement
of the diffraction-ring intensity-profile to discern the difference; a
narrowing of the diffraction rings, as shown in Fig.4.
\begin{figure}[ptb]%
\centering
\includegraphics[
height=2.0003in,
width=3in
]%
{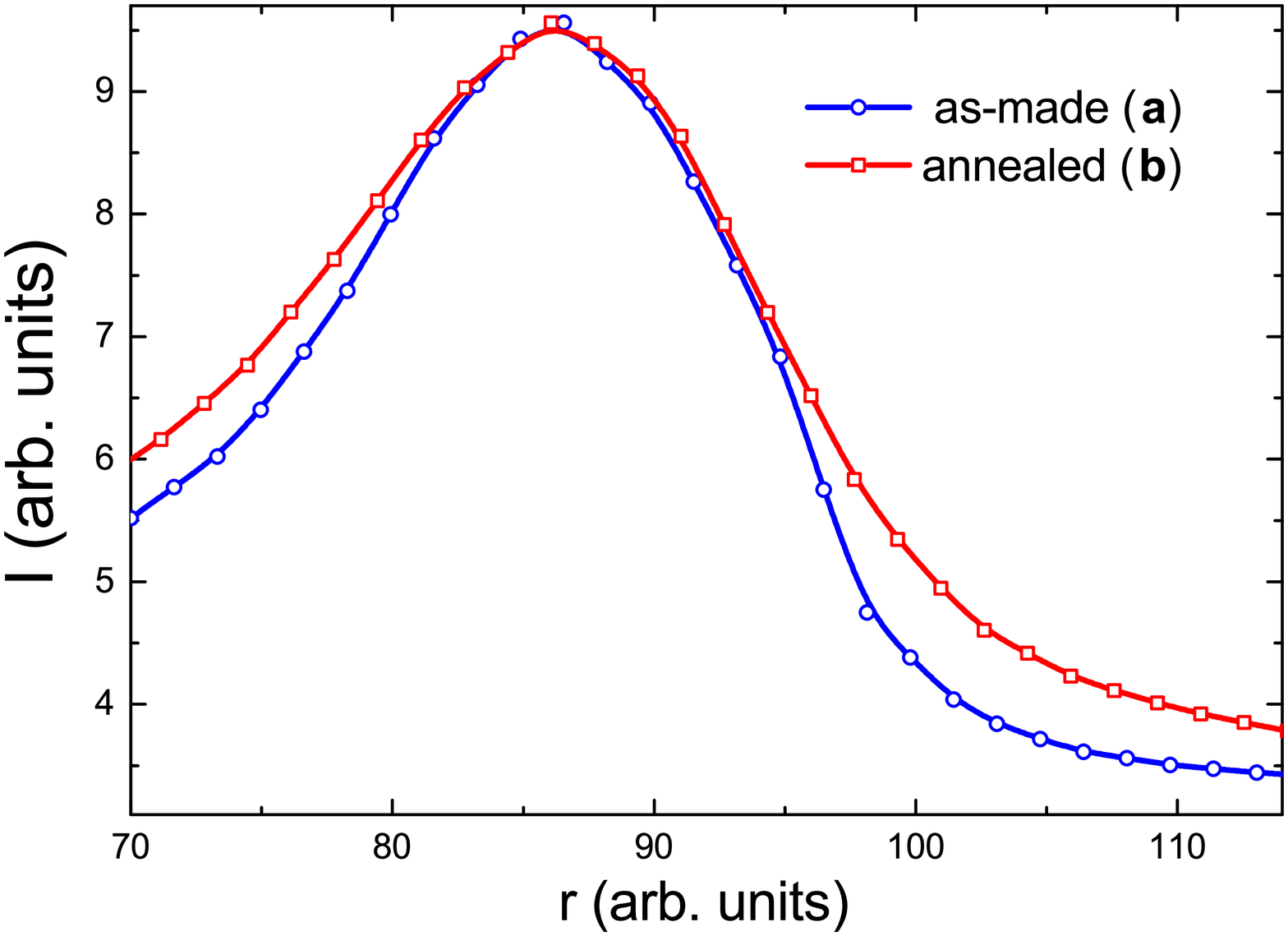}%
\caption{(color online) Intensity profile of the strongest rings in the
diffraction pattern of both (a) and (b) micrographs shown in Fig.3 as function
of distance r from the diffraction center. Data were taken by averaging
line-scans using \textbf{imagej}.}%
\end{figure}
The change in the resistance is essentially due to modified mobility; Hall
effect studies showed only a small change as a result of the annealing process
\cite{46,50}. It was also found that the material volume decreases during the
process which could be detected by measuring the thickness of the film for
example, by x-ray interferometry as demonstrated in Fig.5.%
\begin{figure}[ptb]%
\centering
\includegraphics[
height=2.2061in,
width=3.039in
]%
{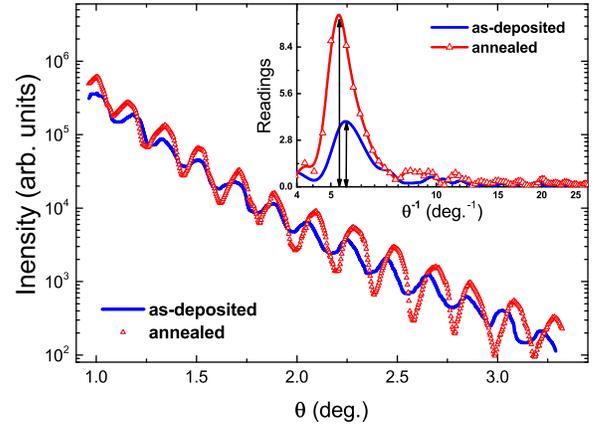}%
\caption{(color online) Reflection-interferometry using a Cu K$\alpha$ x-ray
source (wavelength=1.54\AA ) for the before-after annealing stages of a
similar film as in Fig.3. The inset shows the Fourier transform of the main
plot exposing both the reduced film thickness in the annealed sample and the
enhanced visibility of the interference pattern.}%
\end{figure}

This result was part of an extensive work designed to measure the change in
the optical properties that accompany the thermal-annealing of In$_{\text{x}}%
$O samples \cite{46} with different carrier-concentrations. We bring here
fuller results and interpretation of these data.

The volume-change caused by annealing the sample was reflected in the
absorption versus energy plot as reduction of the optical-gap. This is
illustrated in Fig.6 for one of the studied batches.
\begin{figure}[ptb]%
\centering
\includegraphics[
height=2.2053in,
width=3.039in
]%
{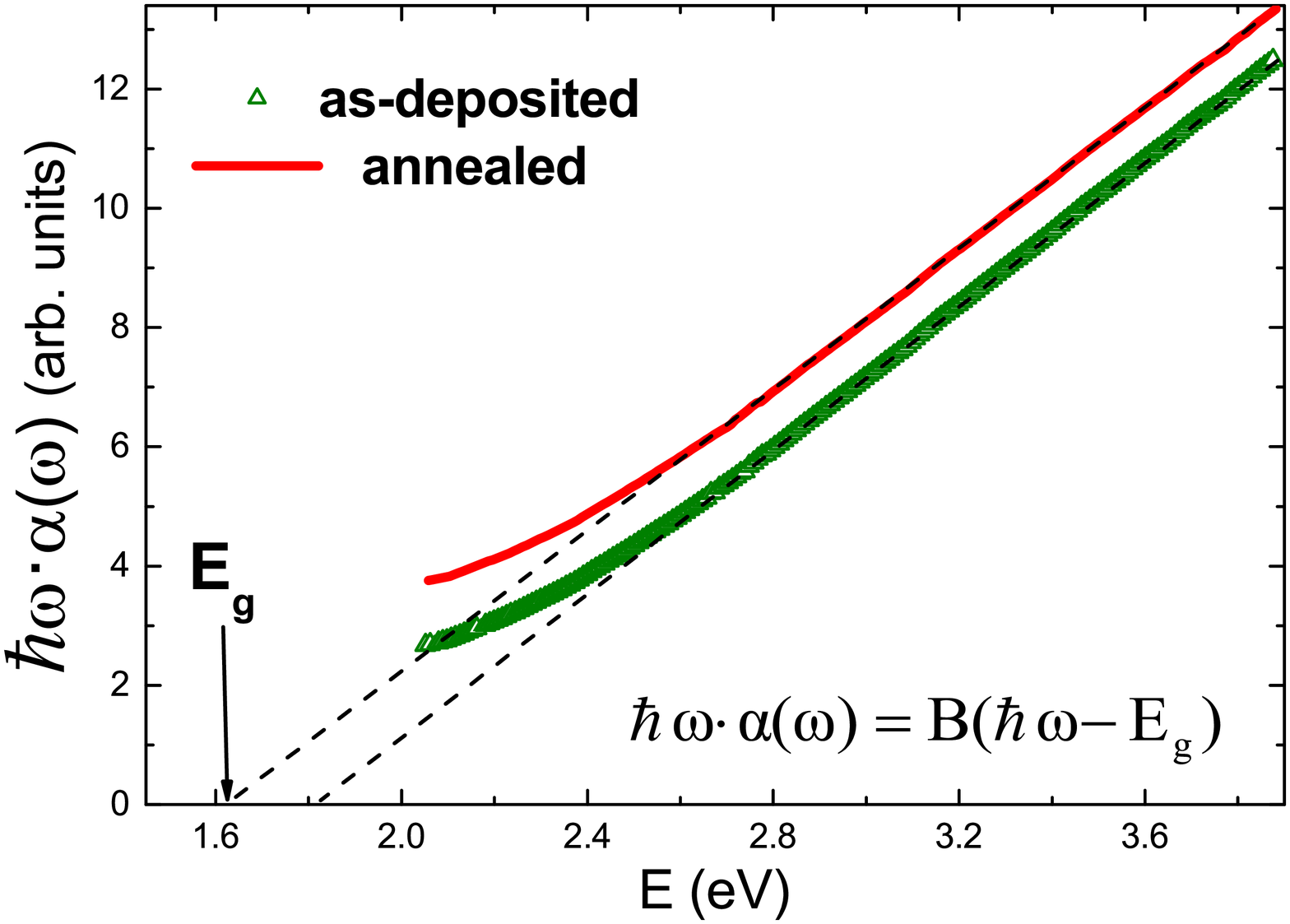}%
\caption{(color online) Absorption versus photon energy for a before-and-after
In$_{\text{x}}$O sample (characterized by carrier-concentration N$\simeq
$8.5x10$^{\text{19}}$cm$^{\text{-3}}$). Dashed lines delineate the respective
values of the energy-gap.}%
\end{figure}

The dependence of absorption coefficient $\alpha$($\omega$) on energy $\omega$
of all our In$_{\text{x}}$O samples obeys the relation:%
\begin{equation}
\alpha(\omega)\hbar\omega=\text{B(}\hbar\omega\text{-E}_{\text{g}}\text{)}%
\end{equation}
A volume-reduction that occurs concomitantly with a smaller optical-gap is
often observed in pressure studies of amorphous materials \cite{51,52,53}. The
increase in wavefunctions overlap due to the reduced volume leads to a wider
band-width and to a modified $\partial$n$/\partial\mu($E$)$. A schematic of
such a change in the conduction-band shape is depicted in Fig.7. Note that
this is depiction is consistent with the observed before-and-after
absorption-curves and with the sum-rule of the number of states. However, it
does not into account a modification in the valence-band and should only be
viewed as a pedagogical aid.%
\begin{figure}[ptb]%
\centering
\includegraphics[
height=1.7149in,
width=3.039in
]%
{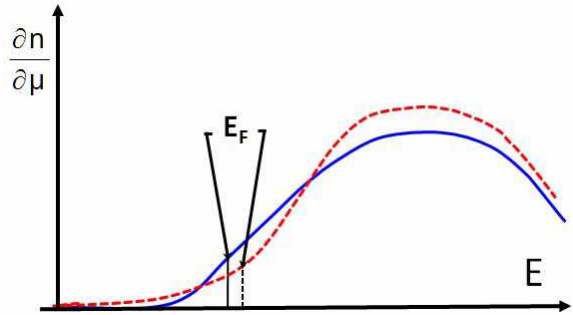}%
\caption{(color online) A schematic depiction of the conduction-band shape for
the as-prepared (dashed line) and annealed (full line) specimen. This assumes
that the change in the valence-band may be neglected. Note that $\partial
$n$/\partial\mu$~at E$_{\text{F}}$ may actually increase during the thermal
annealing due to the combined effect of reduced Lifshitz-tail and mid-band
widening.}%
\end{figure}
This modification of the band shape is a more plausible explanation for the
reduced optical-gap in the thermal-annealing experiments than the effect of
the mobility-edge shift offered in \cite{46}.

The similarity in the effects produced by applying hydrostatic pressure
\cite{51,52,53} and thermally-annealing the sample is not surprising; in both
cases energy is supplied to a metastable system allowing it to cross barriers
associated with its being a structural glass.

Our study include a detailed comparison of the optical properties with
simultaneously measured changes of the disorder parameter k$_{\text{F}}\ell$.
The values for k$_{\text{F}}\ell$ and the Fermi-energy E$_{\text{F}}$ used
here are based on free-electron formulae; E$_{\text{F}}$=$\frac{\hslash
^{\text{2}}(3\pi^{\text{2}}N)^{\text{3/2}}}{2\text{m*}}$ and k$_{\text{F}}%
\ell$=(3$\pi^{\text{2}}$)$^{\text{2/3}}\hbar\sigma_{\text{RT}}$/$e^{\text{2}}%
$\textit{N}$^{\text{1/3}}$ where $\sigma_{\text{RT}}$ is the conductivity
(measured at room-temperature) and m* is the effective mass.

Both, optical and electrical measurements were performed \textit{in-situ} on
thick (900-1100\AA ), effectively 3D films of In$_{\text{x}}$O. Data for the
optical gap E$_{\text{g}}$ were collected at intermediate stages of the
annealing process starting from the as-prepared sample with typically
k$_{\text{F}}\ell\leq$10$^{\text{-3}}$ and ending in k$_{\text{F}}\ell$%
$>$%
1. The process was repeated with films of various carrier-concentrations.
Fig.8 shows the dependence of E$_{\text{g}}$ on k$_{\text{F}}\ell$ for each of
the studied samples.%
\begin{figure}[ptb]%
\centering
\includegraphics[
height=4.1338in,
width=3.039in
]%
{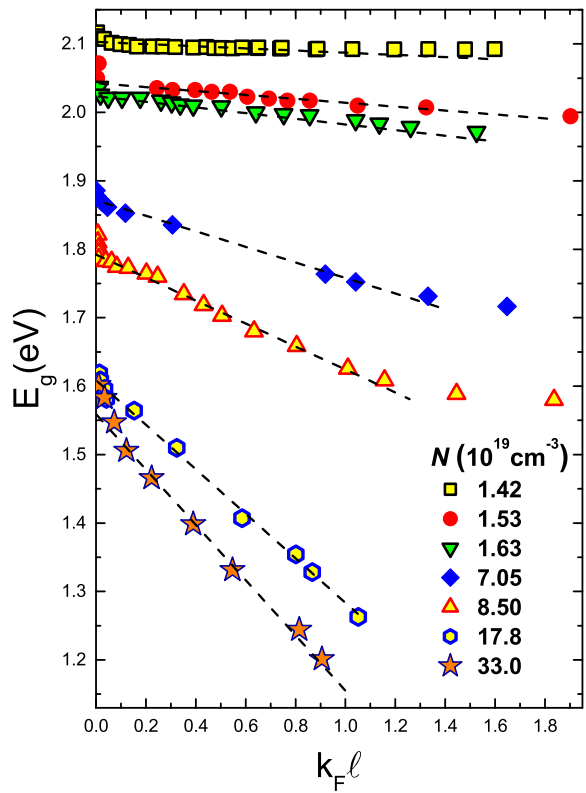}%
\caption{(color online) The dependence of the optical-gap E$_{\text{g}}$ on
the Ioffe-Regel k$_{\text{F}}\ell$ parameter for the seven batches (labeled by
their respective \textit{N}) of In$_{\text{x}}$O samples. Dashed lines are
guides to the eye.}%
\end{figure}

The systematic trend that emerges here is that the optical gap E$_{\text{g}}$
is a monotonous function of k$_{\text{F}}\ell$ or rather, both k$_{\text{F}%
}\ell$ and E$_{\text{g}}$ depend monotonically on the sample volume. Secondly,
for a given change in k$_{\text{F}}\ell$, the concomitant change in
E$_{\text{g}}$ depends on the carrier-concentration \textit{N} of the batch;
$\Delta$E$_{\text{g}}$ is bigger for the batch with the larger \textit{N}. The
change of the refractive index n (judged by the change in the prefactor B in
Eq.3 \cite{54}) during the annealing process showed a similar trend. Fig.9
shows a systematic correlation between E$_{\text{F}}$ and the change in
disorder of n and E$_{\text{g}}$ for each of batches studied (for a
unity-change in k$_{\text{F}}\ell$).%
\begin{figure}[ptb]%
\centering
\includegraphics[
height=2.2753in,
width=3.039in
]%
{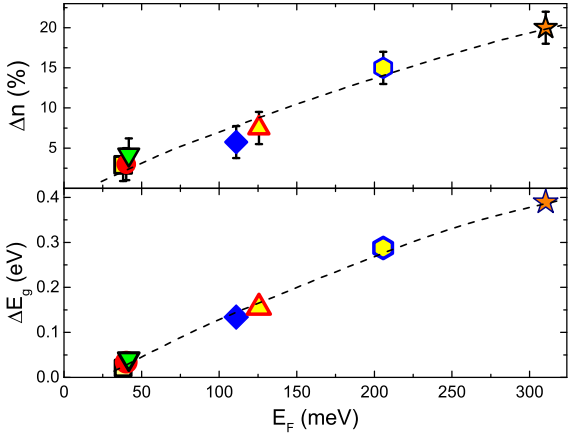}%
\caption{(color online) $\Delta$E$_{\text{g}}$ and $\Delta$n are the
respective changes in the optical-gap E$_{\text{g}}$ and the refractive-index
n when the disorder is changed (by thermal annealing) from k$_{\text{F}}%
\ell\approx$0.02 to k$_{\text{F}}\ell$=1. The data points stand for each of
the batches studied (labeled by symbols corresponding to each of the batches
in Fig.8).}%
\end{figure}

\section{Analysis and Discussion}

Before proceeding It may be a good idea to check whether our assumptions so
far are not at odds with reality. We have focused on In$_{\text{x}}$O because
of available data pertaining to this system with a wide range of
carrier-concentration and disorder. In particular, the metal-insulator
transition has been studied in this material for In$_{\text{x}}$O versions
with \textit{N}$\simeq$10$^{\text{19}}$cm$^{\text{-3}}$ and \textit{N}$\simeq
$10$^{\text{21}}$cm$^{\text{-3}}$. Both versions were found to cross the
transition at a similar k$_{\text{F}}\ell\simeq$0.32$\pm$0.02 \cite{24}. To
localize In$_{\text{x}}$O with E$_{\text{F}}\approx$0.3eV, which is the
highest Fermi-energy in the series studied here, $\mathcal{W}_{\mathcal{C}}$
has to be $\gtrsim$1.9eV (to be consistent with Eq.3). Is such disorder a
viable occurrence in this system?

A main source of disorder in In$_{\text{x}}$O is chemical \cite{24,46}. This
is associated with deviation from stoichiometry; relative to the ionic
compound In$_{\text{2}}$O$_{\text{3-x}}$, there are 5-30\% oxygen vacancies in
In$_{\text{x}}$O \cite{24}. To preserve chemical neutrality some of the indium
atoms must assume a valence of +1 instead of the +3 they have in the
stoichiometric compound. When randomly distributed these valence-fluctuations
form a background potential of e$^{\text{2}}$/r$\approx$5eV (assuming an
average interatomic-separation of r$\approx$3~\AA \ \cite{55}). This type of
disorder is quite prevalent in non-stoichiometric compounds, metallic-oxides,
high-Tc materials etc. The associated disorder $\mathcal{W}$ in these
materials may reach 4-5eV.

In addition to chemical-disorder, there is an \textit{off-diagonal} disorder
in the material that is partially alleviated in the thermal annealing process.
This disorder is related to the distributed nature of the inter-atomic
separation (the random values of inter-atomic separation). The distribution of
inter-particle distances (and thus wavefunction overlap) obviously gets
narrower as the volume decreases and the system approaches the `ideal'
closed-packed amorphous structure. The diminishment of this, off-diagonal
disorder, is clearly reflected in the enhanced visibility of the x-ray
interference pattern of the annealed sample in Fig.4 and also in the electron
diffraction Fig.3 and Fig.4 (albeit less conspicuously).

It has been a controversial issue whether off-diagonal disorder can lead to
localization \cite{56,57,58}. It should be noted that, in the In$_{\text{x}}$O
system, the metal-insulator transition is crossed by the combined effect of
changing the atomic overlap (which in turn affects the density of states as
schematically described in Fig.6) while simultaneously modifying the
off-diagonal-disorder. Localization (at the less-annealed regime) is also
aided by the underlying chemical disorder, and possibly by the Coulomb
interaction \cite{59}.

Two features of the data in Fig.7 and Fig.8 are worth noticing: First, for a
given batch (fixed \textit{N}) there is a systematic and monotonous dependence
of E$_{\text{g}}$ on k$_{\text{F}}\ell$ although the transport measures
essentially just the elements of the CCN while optics is sensitive to the
whole sample. The correlation between the two measurements is indicative of an
underlying common conductance-distribution; being disordered the system is
inherently inhomogeneous but it is so in a generic way (as embodied in the
percolation treatment of hopping conductivity). Inhomogeneity associated with
large thickness or composition variations across the sample will in general
not show the correlation between transport and a bulk measurement. Secondly,
the change in both the optical-gap and the refractive index between
k$_{\text{F}}\ell$=1 and k$_{\text{F}}\ell\simeq$0 exhibits a near-linear
correlation with the Fermi-energy of each batch (Fig.8 and Fig.9).

As noted above, during annealing the sample volume decreases (Fig.5), which
among other things, causes the refractive index to increase. $\Delta$n may
then be used to estimate the associated change in the system energy per
particle $\delta$E. Let us take the largest swing of $\Delta$n in the series
being $\approx$20\% for the sample with E$_{\text{F}}\simeq$300~meV (Fig.8).
The average change of the interparticle-separation in this case is $\approx
$7\% {\small (an estimate consistent with the thickness change upon annealing
measured independently for this sample)} giving $\delta$E$\approx
$0.07x5eV=350~meV. Note that this energy is close to the respective change of
the optical gap for this batch (Fig.8). The near-linearity of $\Delta$n and
$\Delta$E$_{\text{g}}$ with E$_{\text{F}}$ guarantees that proportional
results are obtained for the entire series of samples studied here. This means
that the disorder-energy in the range k$_{\text{F}}\ell$=1 to k$_{\text{F}%
}\ell$=0 changes by no more than E$_{\text{F}}$ of the batch under study.

For the electron-glass phase of In$_{\text{x}}$O however the relevant range of
k$_{\text{F}}\ell$ is only 0%
$<$%
k$_{\text{F}}\ell$%
$<$%
0.32; In$_{\text{x}}$O sample with k$_{\text{F}}\ell\approx$0.02 has a
resistance of $\approx$1G$\Omega$ at 4K \cite{24}, therefore the change in
$\delta$E is even smaller than E$_{\text{F}}$. The entire change of disorder
relevant for the electron-glass phase of In$_{\text{x}}$O can be concisely
summarized as $\delta$E$\lesssim\beta^{\prime}$E$_{\text{F}}$ where
$\beta^{\prime}\approx$0.3. {\small This can be combined with Eq.2 to give a
measure of the disorder }$W_{\text{EG}}$ {\small in the electron-glass regime
(namely, deep in the Anderson-localized phase):}%
\begin{equation}
\mathcal{W}_{\text{EG}}\approx(\beta+\beta^{\prime})\text{E}_{\text{F}}%
=\beta^{\ast}\text{E}_{\text{F}}%
\end{equation}
Most of the disorder in $\mathcal{W}_{\text{EG}}$ is actually due to
$\mathcal{W}_{C}$=$\beta$E$_{\text{F}}$. An order of magnitude estimate for
$\mathcal{W}$ in the diffusive regime may be taken as $\mathcal{W=}\hslash
$/$\tau=$ $\hslash$V$_{\text{F}}$/$\ell$ ($\tau$ is the transport
mean-free-time). With E$_{\text{F}}$=$\hslash^{\text{2}}k_{\text{F}}%
^{\text{2}}$/2m* this yields for the ratio $\mathcal{W}$/E$_{\text{F}}%
$=2/k$_{\text{F}}\ell$. As the transition of In$_{\text{x}}$O is approached
from the diffusive regime k$_{\text{F}}\ell\rightarrow$0.32 and
$\mathcal{W\rightarrow}$ $\mathcal{W}_{C}\approx$6.2E$_{\text{F}}$. Increasing
the disorder above this point, further localizes the system but, as the
resistance is exponential with disorder in this regime, a relatively small
added amount of disorder $\approx$0.3E$_{\text{F}}$ is necessary to get deep (
k$_{\text{F}}\ell\lll$1) into the insulating state.

The near-linear relation between disorder and Fermi-energy established here
for In$_{\text{x}}$O samples is significant; these samples cover more than a
decade range in carrier-concentration and over this range the dynamics is
changing by almost three orders of magnitude \cite{23}. This is a result of
the exponential dependence of relaxation dynamics on $\mathcal{W}_{\text{EG}}$
as is argued next.

Our conjecture is that a relation between $\mathcal{W}$ and E$_{\text{F}}$ of
the form expressed by Eq.5 and with $\beta^{\ast}$of similar magnitude as in
In$_{\text{x}}$O is generally obeyed by Anderson insulators.

If lightly-doped semiconductors follow the same trend, their dynamics should
be much faster than in electron-glasses with \textit{N}$\geq$3x10$^{\text{20}%
}$cm$^{\text{-3}}$. To illustrate, let us use eq.1 on relative basis to
compare the typical tunneling probability for the In$_{\text{x}}$O sample with
\textit{N}$\approx$3.3x10$^{\text{20}}$cm$^{\text{-3}}$ with that of a
lightly-doped semiconductor like Si. Equation.1 with m*$\approx$%
0.3m$_{\text{0}}$, V*$\approx$1.7-1.9eV, and $\mathcal{L}\approx$20\AA \ gives
$\gamma\approx$10$^{\text{-7}}$. The parameters for Si-MOSFET in the Anderson
insulating regime \cite{60} are similar except for V*; given a
carrier-concentration of \textit{N}$\approx$10$^{\text{17}}$cm$^{\text{-3}}$,
typical of the insulating phase in this system \cite{60}, E$_{\text{F}}$ is
$\approx$100 times smaller. Since $\beta^{\ast}$ is likely only smaller when
the Fermi-energy is reduced \cite{48}, the respective $\mathcal{W}$ for Si
should be at least two orders of magnitude smaller than that of the
In$_{\text{x}}$O sample, which means five to six orders of magnitude faster
dynamics (assuming comparable electron-phonon coupling strength and a similar
spectral nature of the disorder). If conductance relaxation in the
In$_{\text{x}}$O may be observed for days, it may not last more than seconds
in lightly-doped semiconductor like Si. With such a relaxation time it would
be practically impossible to observe the memory-dip in field-effect
experiments. It is hard to see how any parameter peculiar to lightly-doped
semiconductor can compensate for their weak disorder. Note that the estimate
made above is just for the \textit{bare} tunneling, it does not take into
account coupling to the environment, which would further increase the
discrepancy between these rates \cite{27}.

A test of these considerations is to instill a disorder of few electron-volts
in a lightly-doped semiconductor: by our conjecture, slow relaxation and
intrinsic electron-glass effects should be observable in this case.
Unfortunately, the measurement of such a sample may turn out to be a tall
order. With such strong disorder ($W$/E$_{\text{F}}\ggg$1) it is doubtful
whether the conductance is measurable even at temperatures that exceed the
Fermi-energy while, like in any other electron-glass one should maintain
T$\ll$E$_{\text{F}}$/k$_{\text{B}}$in such a measurement, which for these
systems means T$\ll$1K. At these temperatures the resistance for a sample with
such $W$ will balloon out of reach.

The difficulty can be demonstrated by reference to an experiment using as a
model, a thin film of In$_{\text{2}}$O$_{\text{3-x}}$, a system where the
resistance can be manipulated over a considerable range. In the experiment,
shown in Fig.10, the room-temperature resistance R$_{\text{RT}}$\ of the
In$_{\text{2}}$O$_{\text{3-x}}$ film is changed by UV-treatment \cite{50}.%
\begin{figure}[ptb]%
\centering
\includegraphics[
height=2.1819in,
width=3.039in
]%
{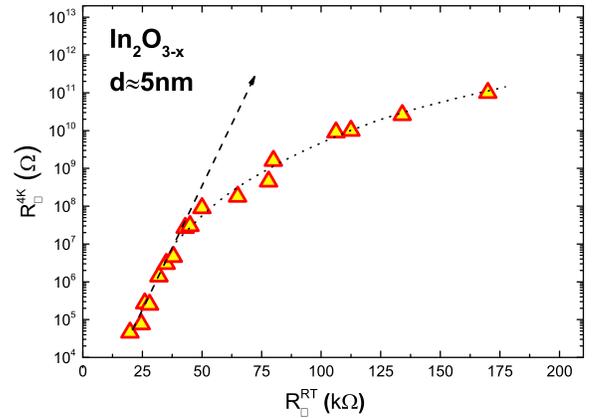}%
\caption{(color online) The sheet resistance of a In$_{\text{2}}%
$O$_{\text{3-x}}$ film at T=4.1K versus its value at room temperature. The
dashed line depicts the expected value of resistance extrapolated along the
line where the room-temperature resistance may be a linear measure of
disorder. The dotted curve is a guide for the eye. }%
\end{figure}
Samples with R$_{\text{RT}}$ spanning the range of $\approx$16-180k$\Omega$
were generated from a single batch in this method. At T$\approx$4.1K these
samples had resistances spanning $\approx$6 orders of magnitude. The value of
R$_{\text{RT}}$ may be taken as a (linear) measure of disorder only when the
sample is in the diffusive regime, R$_{\text{RT}}\lesssim$h/e$^{\text{2}}$.
Strong-localization behavior apparently sets-in for R$_{\text{RT}}\gtrsim
$40k$\Omega$ where the linear relation between disorder and R$_{\text{RT}}$
does not hold. Over the range of R$_{\text{RT}}$=16-32k$\Omega,$ a mere factor
of $\approx$2 in disorder, the resistance at 4.1K changes by $\approx$4 orders
of magnitude. By extrapolation from the regime where R$_{\text{RT}}$ is a
linear measure of disorder, changing it by an order of magnitude would
increase the resistance (at $\approx$4K) by $\approx$20 orders of magnitude,
yielding a resistance that defies current measurements techniques. We suspect
that similar catastrophe would occur upon cranking-up the disorder in
lightly-doped semiconductors.

A more feasible test, is a low-temperature study of a weakly-localized system
(k$_{\text{F}}\ell\gg$1). This can be realized for instance by using a
two-dimensional (2D) film of In$_{\text{2}}$O$_{\text{3-x}}$ which has been
extensively studied in the weakly-localized regime. Being 2D, the system
should crossover to the strongly-localized regime \cite{61}\ at sufficiently
low temperature despite having\textit{ sub-critical }disorder (in the 3D
sense). At sufficiently low temperatures its resistance could be as high as a
sample with k$_{\text{F}}\ell\ll$1 of this material that exhibits glassy
effects at say $\approx$4K \cite{15,22,27}. The logic presented above
anticipates that the weakly-disordered sample would exhibit at best very weak
memory-dip in field-effect experiments (it may not show a null effect due to
occasional regions with potential-fluctuation of large amplitude).

It should be emphasized that by itself, large resistance does not guarantee
electron-glassiness; resistance may result from many factors not related to
the type of disorder discussed here. For example, regions of hard-gaps such as
a series inclusion of band-insulator or isolated islands of superconductivity
in the current-path \cite{24} may exhibit huge resistances without necessarily
showing glassy effects on extended time scales.

Another corollary of the proposed picture is that relaxation-times cannot be
arbitrarily long. This is a result of the maximum value of $\mathcal{W}$
available in reality. Naturally occurring defects in condensed matter systems
have an energy of the order of band-widths, which limits their ability to
Anderson-localize a system. Indeed, to localize a typical metal with
Fermi-energy of $\approx$3-4eV, one has to mix it with another material,
usually a band-insulator, rendering the system granular.

Some of the electron-glass transport features are exhibited by granular
systems in their activated regime. In particular, they show a memory-dip and
slow relaxation \cite{62}. The element of the relaxation in the granular
systems, common with the electron-glass, is re-distribution of charges in
space to lower the `electrostatic' energy of the system. Their dynamics
however is expected to be different than that of Anderson-insulators; charge
re-distribution between different grains is probably controlled by Coulomb
blockade constraints. Charging energies of typical grains could be quite
large, which may lead to extremely long relaxation times. This was
demonstrated by the Bar-Ilan group; in their experiments on several granular
systems it was shown that below few degrees Kelvin dynamics became so slow
that it was impossible to follow the evolution of a memory-dip \cite{63}. The
constraints posed by charging-energies may be partially alleviated by random
fields in the insulating matrix in which the metallic granules are embedded
but then the dynamics is controlled by extrinsic elements. These issues
deserve further study.

In summary, based on the empirical attributes of the electron-glass dynamics,
lead us to conclude that a significant part of the slow relaxation observed in
systems with relatively large carrier-concentration has to do with their much
stronger disorder. The elements of the relaxation process are assumed to be
tunneling events that proceed to ultimately minimize the system energy under
the constraints of disorder and Coulomb interaction. The slowest transitions
naturally occur in the dead-wood regions of the system that in the hopping
regime occupy most of the system volume. Slow `fluctuators' are abundant in
these regions even for a rather short tunneling distance once the disorder is
appreciable. An estimate of the disorder-energy $\mathcal{W}$ for a realistic
sample that shows electron-glass properties was made using In$_{\text{x}}$O as
a model system. This was based on measurements of the metal-insulator
transition made on these compounds, free-electron concepts, and on optical
data relevant for their strongly-disordered regime. It was shown that
$\mathcal{W}$ is proportional to the Fermi-energy of the Anderson-insulating
system for the entire range of disorder relevant for the electron-glass
measurements. All other things being equal, and given the exponential
dependence of the tunneling rate on $\mathcal{W}$, one expects
energy-relaxation in lightly-doped semiconductors where E$_{\text{F}}$ is two
orders of magnitude smaller, to be many orders of magnitude faster as indeed
established experimentally \cite{25}.

There is still the challenge of accounting for the protracted relaxation-times
observed in the experiments. This is a problem common to all glasses but the
flexibility of experimenting with electron-glasses seem to offer more scope
for progress. From the theory point of view the problem has proved to be
difficult; Even when a complete knowledge of the disorder is at hand there are
other pieces of the puzzle that need careful elaboration. In particular,
coupling to the environment may play a role in further slowing-down
transition-rates. Environmental degrees of freedom that follow adiabatically
the tunneling object would modify the transition rate through mass enhancement
(polaronic effects). More generally, coupling to the environment will suppress
tunneling due to the Anderson orthogonality catastrophe (AOC) \cite{27}.
Recent work focused on new aspects of the AOC (originally conceived for clean
systems \cite{64}) in the strongly-localized and interacting regime
\cite{65,66}. However, a model incorporating coupling to the environment for a
medium lacking screening (like an Anderson-insulator), is more pertinent for
our experiments. To our knowledge, this aspect has not been adequately
addressed by any work so far. Other effects that probably contribute to slow
relaxation are hierarchical constraints (the `domino' effect being a special
case), and correlated many-electron transitions. Fundamental questions related
to these scenarios are yet to be resolved. However, to understand why a
memory-dip is not likely to be observed in lightly-doped semiconductors by
transport measurements, it may suffice to consider just the role of disorder.

\begin{acknowledgments}
The author gratefully acknowledges illuminating discussions with Eldad
Bettelheim and Ori Grossman and for sharing insight based on their numerical
study. This research has been supported by a grant No 1126/12 administered by
the Israel Academy for Sciences and Humanities.
\end{acknowledgments}


\begin{thebibliography}{99}
\bibitem {17}The field-effect, much like the Hall-effect is usually conceived
as a thermodynamic measurement and then it is $\partial$n$/\partial\mu$~that
it is probed not the single-particle density of states. The situation is
different when the gate is scanned with a rate that is faster than the
relaxation time of the electronic system.
\bibitem {43}This trend has been consistently observed in each and every
system that exhibited the memory-dip.
\end{thebibliography}
\end{document}